\documentclass[final]{IEEEtran}

\usepackage{cite}
\usepackage{color}
\usepackage{threeparttable}
\usepackage[pdftex]{graphicx}
\usepackage{epstopdf}
\usepackage{picinpar}
\usepackage[cmex10]{amsmath}
\usepackage{amsmath,amsfonts,amssymb}
\usepackage{subfigure}
\usepackage{changepage}
\usepackage{algorithm}
\usepackage{caption}
\usepackage{algpseudocode}
\usepackage{stfloats}
\usepackage{bm}
\usepackage{amsthm}
\usepackage{enumerate}
\usepackage{multirow}
\usepackage{booktabs}
\usepackage{url}

\allowdisplaybreaks
\begin{document}
\newtheorem{lemma}{Lemma}
\newtheorem{corollary}{Corollary}
\newtheorem{theorem}{Theorem}
\newtheorem{proposition}{Proposition}
\newtheorem{definition}{Definition}
\newcommand{\e}{\begin{equation}}
\newcommand{\ee}{\end{equation}}
\newcommand{\eqn}{\begin{eqnarray}}
\newcommand{\eeqn}{\end{eqnarray}}
%  调整公式间距
\newenvironment{shrinkeq}[1]
{ \bgroup
\addtolength\abovedisplayshortskip{#1}
\addtolength\abovedisplayskip{#1}
\addtolength\belowdisplayshortskip{#1}
\addtolength\belowdisplayskip{#1}}
{\egroup\ignorespacesafterend}
\title{The Road to Industry 4.0 and Beyond:\\
A Communications-, Information-, and Operation Technology Collaboration Perspective}

\author{Ziwei Wan, Zhen Gao, Marco Di Renzo, \IEEEmembership{Fellow,~IEEE}, and Lajos Hanzo, \IEEEmembership{Life Fellow,~IEEE}

\vspace{-5mm}

%\thanks{This work was supported in part by the Natural Science Foundation of China (NSFC) under Grant 62088101, Grant 62071044, and Grant 61827901.}
%\thanks{L. Hanzo would like to acknowledge the financial support of the Engineering and Physical Sciences Research Council projects EP/P034284/1 and EP/P003990/1 (COALESCE) as well as of the European Research Council's Advanced Fellow Grant QuantCom (Grant No. 789028)}
%\thanks{The work of M. Di Renzo was supported in part by the European Commission through the H2020 ARIADNE project under grant agreement number 871464 and through the H2020 RISE-6G project under grant agreement number 101017011.}

%\thanks{Z. Wan, and Z. Gao are with School of Information and Electronics, Beijing Institute of Technology, Beijing 100081, China. (e-mail: \{ziweiwan, gaozhen16\}@bit.edu.cn).}
%\thanks{M. Di Renzo is with Universit\'e Paris-Saclay, CNRS and CentraleSup\'elec, Laboratoire des Signaux et Syst\`emes,  Gif-sur-Yvette, France. (e-mail: marco.direnzo@centralesupelec.fr).}

}

\maketitle

\begin{abstract}
The fourth industrial revolution, i.e., Industry 4.0, is evolving all around the globe. 
In this article, we introduce the landscape of Industry 4.0 and beyond empowered by the seamless collaboration of communication technology (CT), information technology (IT), and operation technology (OT), i.e., CIOT collaboration. Specifically, CIOT collaboration is regarded as a main improvement of Industry 4.0 compared to the previous industrial revolutions.
We commence by reviewing the previous three industrial revolutions and we argue that the key feature of Industry 4.0 is the CIOT collaboration.
More particularly, CT domain supports ubiquitous connectivity of the industrial elements and further bridges the physical world and the cyber world, which is a pivotal prerequisite.
Then, we present the potential impacts of CIOT collaboration on typical industrial use cases with the objective of creating a more intelligent and human-friendly industry.
Furthermore, the technical challenges of paving the way for the CIOT collaboration with an emphasis on the CT domain are discussed.
Finally, we shed light on a roadmap for Industry 4.0 and beyond.
The salient steps to be taken in the future CIOT collaboration are highlighted, which may be expected to expedite the paradigm shift towards the next industrial revolution.
\end{abstract}

\IEEEpeerreviewmaketitle
\section{Introduction}
Historically, the {\bf Industrial Revolution} took place based on the development of science and technology, substantially liberating human productivity at the time. By the time of writing, three industrial revolutions have taken place, which were fuelled by steam- and electric power, as well as by information technology. 
Currently, the fourth industrial revolution, a.k.a., {\it Industry 4.0}, is afoot across the globe. Industry 4.0 is expected to usher in a new era of intelligent, human-friendly industry, and create an unprecedented global market \cite{GE}.

Initially proposed in Germany in 2011 as a part of the German economic policy \cite{IEM}, the concept of Industry 4.0 has been generalized over the past few years as the convergence and application of various advanced techniques \cite{Book}, such as the Internet of Things (IoT), digital twin, big data, artificial intelligence (AI), XR (including {\it virtual reality, augmented reality, and mixed reality}), etc. {\color{black}More recently, these advanced techniques have been amalgamated and expanded as {\it the Metaverse} \cite{Metaverse}, a promising paradigm shift for not only Industry 4.0 but way beyond.} It has attracted many countries' attention at both the enterprise and the governmental levels.
%The main themes of Industry 4.0 include but are not limited to:
%\begin{itemize}
%{\item {\it Smart Factory,} which focuses on the research of intelligent production systems and distributed networks to realize a high-efficiency, environmentally-friendly factory.
%}
%{\item {\it Smart Manufacturing,} which involves the production process management for enterprises, man-machine interaction, and the application of three-dimensional (3D) printing technology in the industrial production process.
%}
%{\item {\it Smart Logistics,} which supports automatic operation and high-accuracy optimization management, while reducing the costs and consumption of natural as well as social resources in the cargo transportation process.
%}
%\end{itemize}

\begin{figure*}[!tp]
%\vspace{-3mm}
\captionsetup{singlelinecheck = off, font={footnotesize}, name = {Fig.}, labelsep = period}
\centering
\includegraphics[width=5.5in]{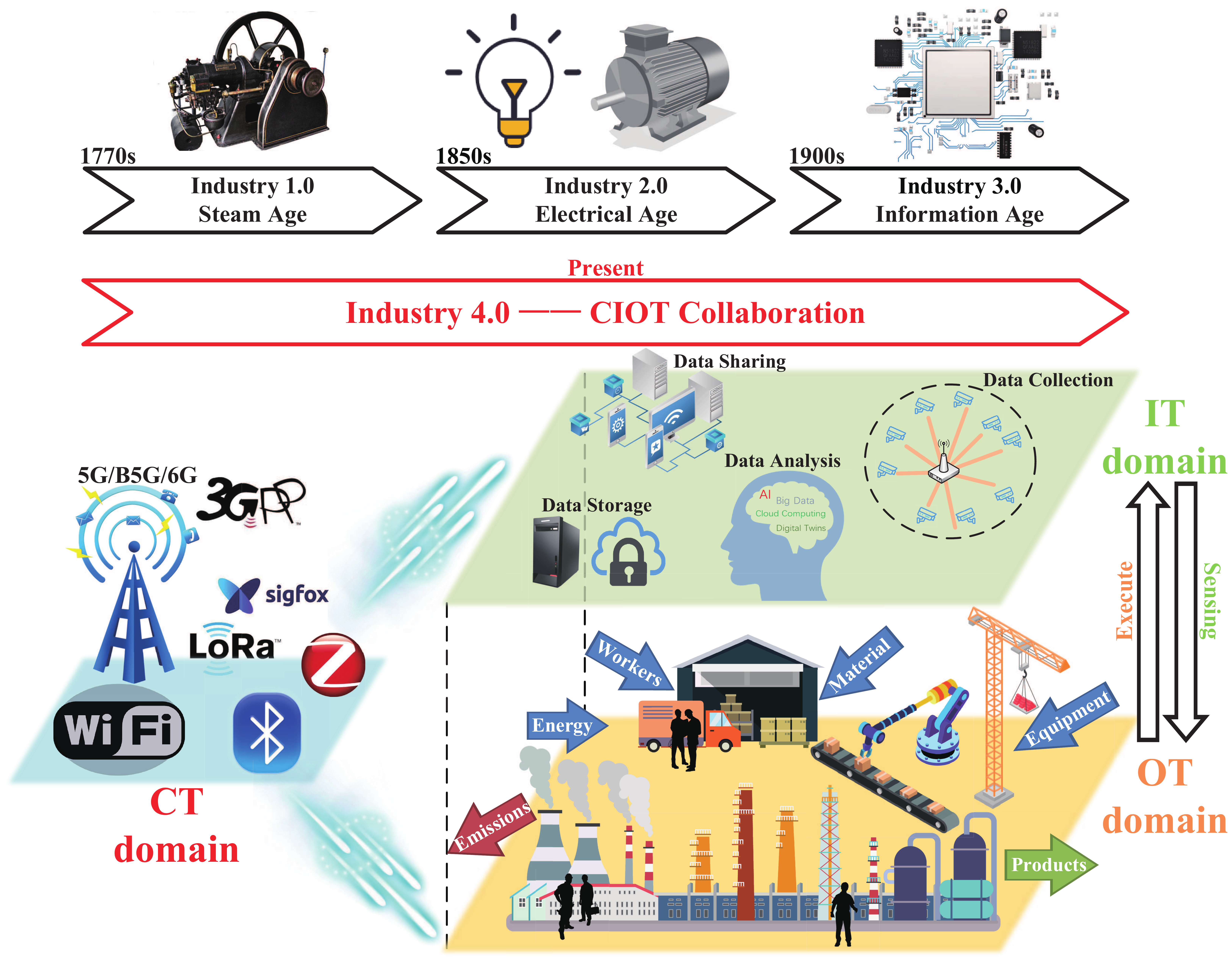}
\caption{The procedures and representatives of the previous three industrial revolutions, and the cooperation mechanism in Industry 4.0.}
\label{F1}
\end{figure*}

The core of Industry 4.0 lies in {\it interconnection}. Industry 4.0 is also closely related to the industrial IoT (IIoT) \cite{TII,Survey}.
Communication technology (CT) forms an indispensable part of Industry 4.0, since it intrinsically amalgamates the sophisticated operation technology (OT) and information technology (IT) domains in support of the efficient IIoT, as shown in Fig.~\ref{F1}.
The performance of CT directly determines the stability, continuity, and flexibility of the global industry.

Nevertheless, establishing a reliable CT domain for industrial applications is a non-trivial task, mainly due to the large-scale connectivity requirements in the face of hostile propagation.
Cisco predicted that by 2023, a significant fraction (about 50\%, i.e., 14.7 billion) of connected devices and connections will come from machine-to-machine (M2M) interaction \cite{Cisco}, which is the common form in industrial scenarios. 
%This stunning increase of connectivities, followed by the transmission of huge amounts of data, imposes challenges on the CT design in Industry 4.0.
{Things are more urgent for Industry 4.0 and beyond, which requires not only extremely demanding connectivity, but also time sensitivity and security attributes.
Fortunately, the rapid development of CTs, including fifth-generation (5G) and even 6G systems, provides effective solutions for Industry 4.0.}
For instance, recalling the three basic scenarios in 5G, i.e., the enhanced Mobile Broadband (eMBB), Ultra Reliable Low Latency Communications (URLLC) and massive Machine Type Communications (mMTC), it is expected that the proliferation of 5G techniques will support massive low-latency connectivity, high-rate data transmission, real-time remote control, and high-safety information exchange in industrial applications. 
Industry 4.0 critically hinges on state-of-the-art CTs.

This article provides an overview of Industry 4.0 and beyond (Industry 4.0+) based on a CT-IT-OT (CIOT) collaboration perspective. 
%After briefly comparing the fourth industrial revolution to the previous ones, we postulate that the CT-IT-OT (CIOT) collaboration is the foundation of Industry 4.0, where CT domain supporting ubiquitous connectivity for massive industrial elements is one of the corner-stones. We also discuss how the CIOT collaboration will reshape the use cases in future industrial applications. 
%More particularly, the technical challenges on the road to CIOT collaboration in Industry 4.0 are summarized, and the CIOT-related research directions of CT domain are introduced. Moreover, we speculate on the ambitious vision of Industry 4.0 and beyond empowered by the inspiring advances in the CIOT domains, which pave the way for the associated scientific research and technical development.
{\color{black}
In particular, the contributions of this article are threefold:
\begin{itemize}
{\item After briefly comparing the fourth industrial revolution to the previous ones, we postulate that the CIOT collaboration is the foundation of Industry 4.0
We argue by introducing some critical use cases that the core of Industry 4.0 and beyond lies in the interconnection of a massive number of the existing industrial elements, leading to CIOT collaboration.
}
{\item The technical challenges on the road to CIOT collaboration in Industry 4.0 are summarized, and the CIOT-related research directions of the CT domain are introduced, mainly including massive access to the IIoT, information transmission and network management.
Some case studies are also conducted to show how the advanced CT domain reshapes the industrial activities.
}
{\item We speculate on the ambitious vision of Industry 4.0+ empowered by a suite of inspiring recent advances, such as the industrial Internet of Senses (IIoS), energy-harvesting networks (EHN), data-driven CIOT (DCIOT) collaboration using AI, and the industrial Metaverse, which further pave the way for the associated scientific research. These are also juxtaposed to the corresponding limitations to be tackled for providing guidelines for the community's future research.
}
\end{itemize}
}

\section{Overview of Industrial Revolutions and Development Roadmap of Industry 4.0}
In this section, we first highlight the approach of the previous three industrial revolutions, and then provide a blueprint for the evolving Industry 4.0, wherein the CIOT collaboration is a key feature. In particular, we focus on how CIOT-related CT domain will play a dominant role in Industry 4.0.

{\color{black}
\subsection{Industry 1.0 to 3.0}
The {\it first} industrial revolution ({\it Industry 1.0}) was triggered by
the first ever commercially successful engine conceived by T. Newcomen
in 1712 that replaced 'muscle-power' by uninterrupted
'machine-power'. This was then further improved by J. Watt in 1764.
The foundations of the {\it second} industrial revolution ({\it
  Industry 2.0}) were laid by M. Faraday, who formulated the
theoretical and practical basis of harnessing electric
power~\cite{Book}.  In contrast to the above two revolutions, which
were characterized by the different types of energy sources, the {\it
  third} industrial revolution ({\it Industry 3.0}) focused on
improving industrial production by jointly using miniaturized
electronics and novel IT, paving the way for industrial
digitization. In the same era, modern CT has been pioneered from a
practical perspective by G. Marconi in the 1900s and from a
theoretical perspective by C. Shannon in the 1940s, providing
compelling opportunities for the digital electronic industries.  All
the previous three industrial revolutions had a huge impact both on
today's industrial production, as well as on wealth-creation, and on
people's quality of life.
%From then on, manpower could be replaced by mechanical power in industrial production, inaugurating the Steam Age,  during which multi-functional machines were invented.
%Electricity was used in industrial production as a new energy source that supplemented and gradually replaced the steam engine.
%The ensuing industrial electrification heralded the Electrical Age, which was considered to be the most essential engineering achievement of the 20th century.
%Under these trends, the inventions of general-purpose computers (1940s), Internet (1960s), smart phones (2000s) and so on, have all made a huge impact on the industrial production, as well as people's lives today.
}
\begin{figure*}[!tp]
%\vspace{-3mm}
\captionsetup{singlelinecheck = off, font={footnotesize}, name = {Fig.}, labelsep = period}
\centering
\includegraphics[width=5in]{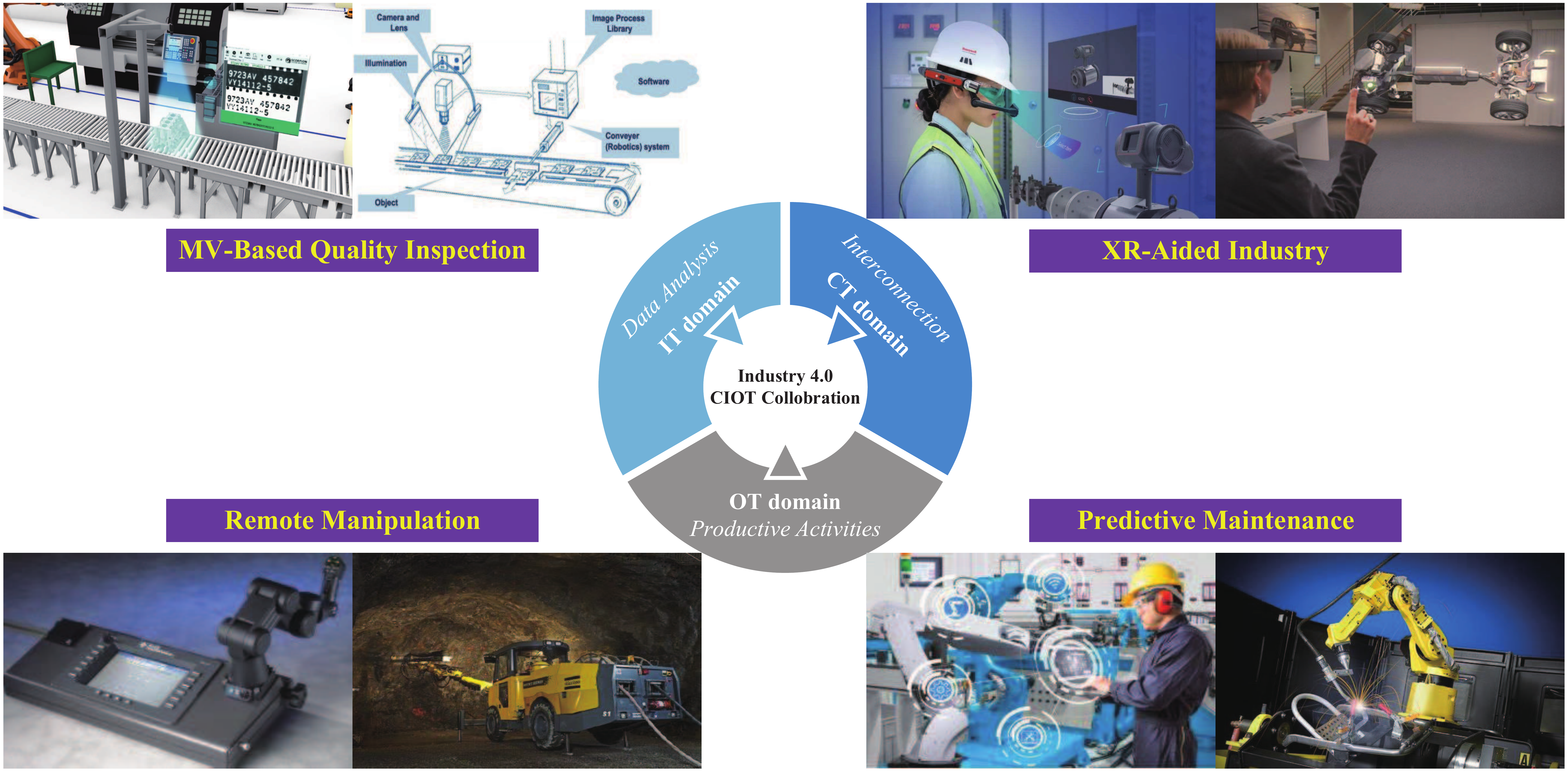}
%\vspace{-1mm}
\caption{Potential use cases of Industry 4.0 empowered by the CIOT collaboration.}
\label{F2}
\end{figure*}

\subsection{Industry 4.0 Based on CIOT Collaboration}
The {\it fourth} industrial revolution, i.e., {\it Industry 4.0}, hinges on the close collaboration of all three CIOT domains in the process of industrial production, as shown in
Fig. 1.
The specific missions of the CIOT domains are as follows:
\begin{itemize}
{\item {\it The OT domain} encompasses all the industrial elements, which are directly responsible for the operation and maintenance of factories or businesses. It consists of workers, equipment, materials, energy, products, emissions, {etc.}, in all the industrial activities.
}
{\item {\it The IT domain} is the platform that realizes the unified collection, storage, and analysis of the data collected by workers or sensor networks from the OT domain. It is empowered by advanced cloud computing, digital twin, AI, etc.
}
{\item \color{black} {\it The CT domain} is represented by various wired/wireless and long-/short-distance communication standards and technologies, such as Wi-Fi, Bluetooth, Zigbee, LoRa, Sigfox, NB-IoT, and the cellular networks{\footnote{\color{black}The length-limitation of the magazine paper precludes the detailed introduction to these standards or techniques. Interested readers are referred to the rich set of related publications.}}. They support networking at a given level of service quality assurance for the intelligent control of industries.
}
\end{itemize}

Based on the processes of Industry 1.0 to 3.0 discussed in the previous sub-section, we can see that although the OT, IT and CT have already been respectively developed for industrial applications,
%In Industry 1.0 and 2.0, the upgrade of energy and machine types reflected the improvement in the OT domain. In the third revolution, the IT domain and the CT domain were introduced, {and their intrinsic amalgam exemplified by} {\it embedded systems} has emerged. However, during the third revolution, the CT domain generally acted as communication tools for humans in human-type communication, rather than for industrial elements, such as machines or sensors in machine-type communication. 
the CIOT collaboration has not been recognized as a research direction until Industry 4.0 has been launched.
As indicated by the terminology, the CIOT collaboration aims for the organic integration of all three CIOT domains, realizing the interaction between the physical world (OT domain) and the digital world (IT domain) via the communication networks (CT domain).
As Fig. 1 indicates, 
%the OT domain executes the specific operations required by the industrial activities according to the strategies gleaned from the IT domain, and the IT domain senses all events in the OT domain by performing data collection, storage, sharing, and analysis in order to optimize the current industrial policy and provide more efficient strategies for the OT domain. 
in the ``sensing'' and ``execute'' process between the IT domain and the OT domain, the CT domain plays a key role, seamlessly amalgamating the physical and the digital worlds. The CT domain becomes so important for the industrial revolution that the ``IIoT'', where the CTs constitute the foundations, is usually regarded as the indispensable component of Industry 4.0 \cite{TII,Survey}.

%Here we would like to stress that the CIOT collaboration is the tangible visualization of the popular concept of CPS, where the OT domain and the IT domain correspond to the {physical world} and the {cyber world} in CPS, respectively.
%Moreover, the role of the CT domain is demonstrated more explicitly in the description of the CIOT collaboration, which provides better understandings how physical world and cyber world in CPS cooperate with each other in industrial scenarios.

\begin{figure*}[t]
%\vspace{-3mm}
\captionsetup{singlelinecheck = off, font={footnotesize}, name = {Fig.}, labelsep = period}
\centering	
\includegraphics[width=6.5in]{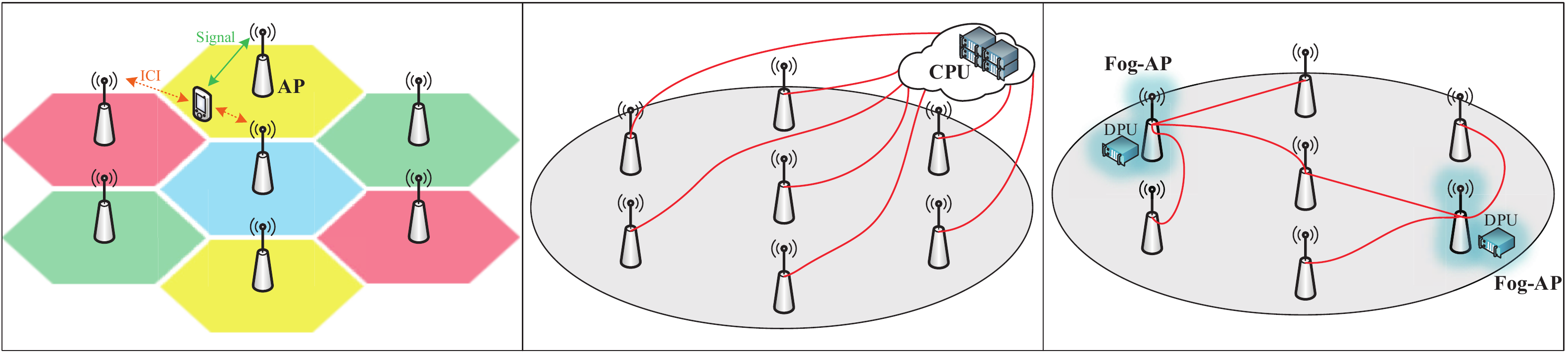}
%\vspace{-3mm}
\caption{Three processing paradigms for multi-AP grant-free massive access: (from left to right) non-cooperate architecture, cloud computing architecture, and edge (fog) computing architecture.}
\label{F3}
\end{figure*}

\subsection{Impact on Use Cases}
Below we discuss the potential use cases in Industry 4.0 empowered by the CIOT collaboration as shown in Fig.~\ref{F2}. We will further demonstrate below that the CT domain is an integral part of hitherto unexplored promising industrial applications of the near future.

{\bf 1) Machine Vision (MV)-Based Quality Inspection}. {\color{black}The vision information of the products on the production line is collected through high-definition industrial cameras, and then it is transmitted through the uplink by relying on the CT domain.}
The CT domain ensures high-integrity, low-latency data transmission.
Also, the AI-based MV algorithms {running} on the IT platform will process, analyse, and understand the images received. On this basis, the machines in the IT domain can identify the targets and detect the quality of products on the production line in near-real-time. Compared to the traditional manual quality inspection, the MV-based quality inspection is of higher sensitivity, higher precision, and higher efficiency, which helps reduce the human cost in the manufacturing process.

{\bf 2) XR-Aided Industry}. Based on {\it virtual reality, augmented reality, and mixed reality} (XR) technologies, the applications of remote assistance, guidance, etc. can be carried out in support of the employees, who learn from each other, work together, and interact with each other. {\color{black}For example, the scenes captured in the production line can be directly passed to the product designer in the form of XR, so that he/she can provide intuitive guidance on improving the production process via reliable communication networks provided by the CT domain.} More importantly, XR-based techniques are capable of reminding the workers operating the assembly lines of precautions or raising the alarm in the case of emergencies. The fully-fledged Industry 4.0 will provide transcoding, 3D reconstruction, object recognition, content management, and other sophisticated XR-based operations.

{\bf 3) Remote Manipulation}. Delicate remote manipulation is of great importance in many industrial scenarios, such as post-disaster reconstruction, replacing manual labor and thus protecting workers from danger.
{\color{black} }
%Generally, managers remotely operate the equipment in the central control room, and adjust the strategy based on the data collected and fed back by numerous sensors.
%More importantly, the equipment should make simple decisions automatically if necessary (e.g., avoiding the obstacles that cannot be seen by managers). 
{\color{black}This application requires the ubiquitous connections anytime and anywhere, which is hard to fulfilled by the conventional terrestrial cellular networks.
As a remedy, the non-terrestrial network (NTN) will serve as a building block for remote manipulation in the dangerous or unpopulated areas. The utilization of NTN will be further elaborated in Section IV-A.
%to offload elementary learning and data processing abilities to the OT domain, enabling {\it machine intelligence} \cite{Pareto}.
Remote manipulation in Industry 4.0 has tight delay specifications (usually milliseconds level \cite{Access}), hence ultra-reliable and low-latency communications are pivotal prerequisites.
}

{\bf 4) Predictive Maintenance}. {\color{black}Predictive maintenance system can capture some precursors of machine failure by leveraging IoT sensor networks, machine learning algorithms, and big data analysis.}
This allows machine owners/operators to perform maintenance in advance of its due date, and thus avoids the sudden failure of the production line, which may cause immeasurable loss.
%, namely before the machinery fails or the materials get damaged.
%This application avoids the sudden failure of the production line, which may cause immeasurable loss.
{\color{black}According to \cite{DT}, predictions on the
remaining useful life span of the components may be carried out by the digital twin, which constitutes a tangible example of the proposed CIOT collaboration.}
{\color{black}Naturally, unless the tight communication specifications are satisfied, it is impossible for managers to remotely monitor the status of machines, and hence we cannot benefit from predictive maintenance.}

\section{Research Directions for the CT Domain}
Despite the appealing visions of this emerging industrial revolution, the existing schemes cannot satisfy the tight requirements of the so-called vertical industries in Industry 4.0.
Based on the discussions of use cases, we readily see that many technical challenges accrue from the CT domain, which again indicates that the CT domain plays a pivotal role in enabling Industry 4.0. In the rest of this section, we highlight the research directions for the CT domain by reviewing three main aspects of the recently developed CTs, i.e. access paradigm, information transmission, and network management, by discussing their potential applications in Industry 4.0.

\subsection{Massive Access for IIoT}
%Improving the levels of intelligence and efficiency in Industry 4.0 is expected to increase the number of sensors and edge processing elements integrated into the machines or vehicles (i.e., {IIoT scenario}), which leads to new challenges for the radio access network (RAN).
The increasing number of sensors and edge processing elements integrated into the IIoT scenario leads to new challenges for the radio access network (RAN).
Massive access technology is the backbone of typical IIoT applications operating under Industry 4.0.
In contrast to conventional grant-based access schemes, grant-free massive access schemes allow multiple IIoT nodes to share the same physical time and frequency resources to send their pilot sequences, based on which the subsequent active terminal detection is performed at the access points (APs). As a benefit, reduced latency can be achieved \cite{MKe}.
Furthermore, due to the sporadic tele-traffic generation of IoT applications, only a small portion of IIoT nodes is simultaneously active.
This sparse activity can be effectively exploited by advanced compressive sensing (CS) algorithms, and solves the active node detection problem with the aid of non-orthogonal pilot sequences.

\begin{figure}[b]
%\vspace{-3mm}
\captionsetup{singlelinecheck = off, font={footnotesize}, name = {Fig.}, labelsep = period}
\centering	
\includegraphics[width=2.5in]{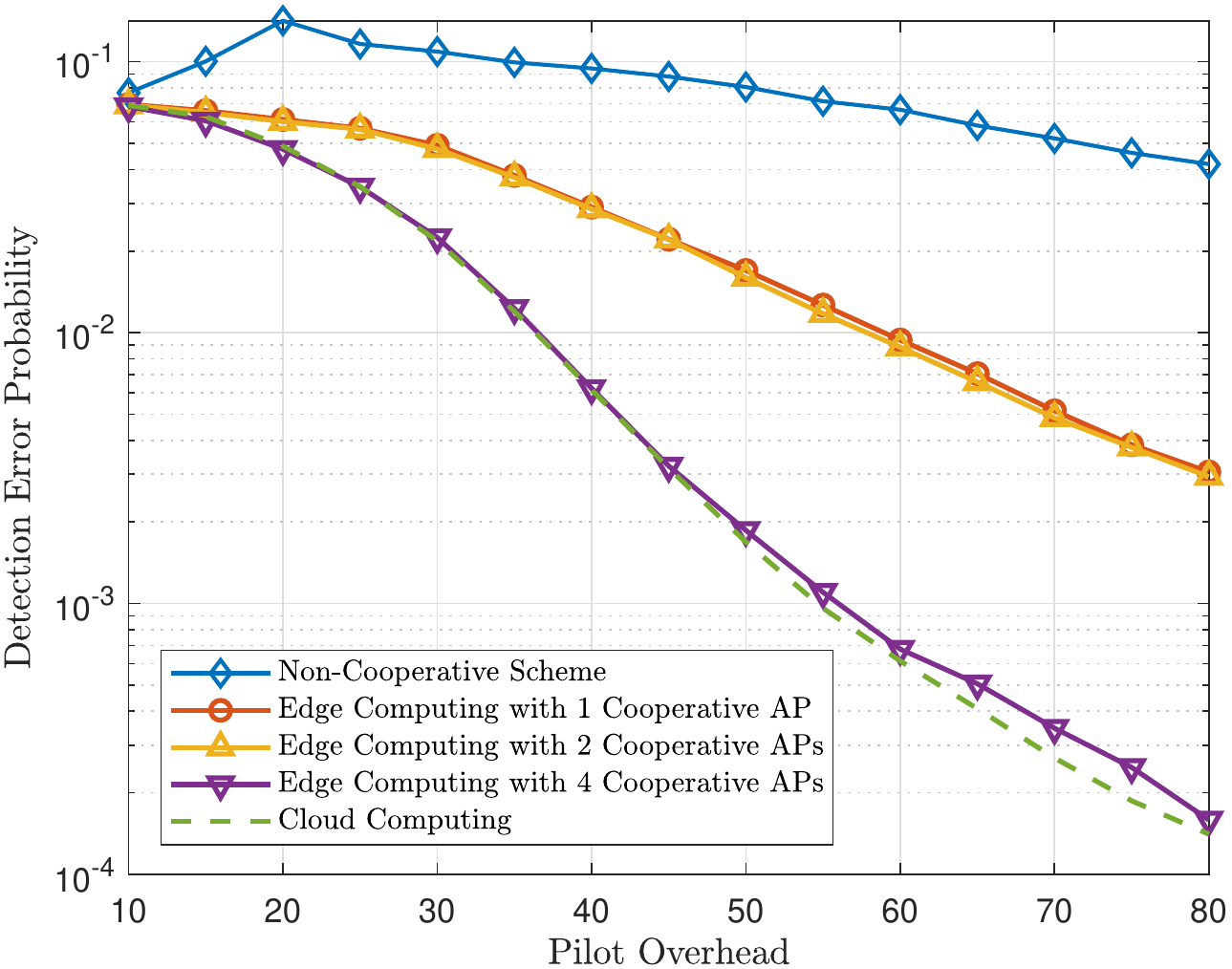}
%\vspace{-3mm}
\caption{Performances of different multi-AP grant-free massive access paradigms. 
%7 APs serve 140 active terminals out of 2800 terminals.
A CS-based algorithm in \cite{MKe} is considered.}
\label{F4}
\end{figure}

The IIoT nodes of large-scale industrial activities in Industry 4.0 are expected to be distributed across a vast area. The APs should cooperate to offer improved coverage and to reduce the transmit power of energy-constrained nodes.
In Fig.~\ref{F3}, three different processing paradigms are considered for multi-AP grant-free massive access \cite{MKe}. In the non-cooperative architecture of Fig.~\ref{F3}, each IoT node is attached to a single AP, where reduced-dimensional signal processing can be conducted at the cost of potentially severe inter-cell interferences (ICI). By contrast, the cloud computing and edge computing architectures of Fig.~\ref{F3} allow each AP to serve all the nearby nodes, and these APs are connected either to a single or multiple processing units for joint signal processing, which blurs the traditional concept of ``cells'' and thus avoids ICI by constructing user-centric clusters relying on fair load-balancing. Compared to cloud computing, where only a single central processing unit (CPU) is responsible for all APs, edge computing offloads the signal processing tasks to multiple distributed processing units (DPUs) deployed at some of the APs (namely, fog-APs), which alleviates the burden imposed both on the backhaul links and on the CPU.
Fig.~\ref{F4} characterizes the performances of these three paradigms.
We observe that with the advent of AP-cooperation, ICI-free cloud computing and edge computing become feasible, which improves the active terminal detection probability.
Furthermore, by increasing the number of cooperating APs, the performance of edge computing becomes capable of approaching that of cloud computing, despite its reduced latency. Hence, edge computing is a more appealing scheme for Industry 4.0 in support of delay-sensitive applications.

\subsection{Wireless Information Transmission}
Due to the explosive growth of connected industrial devices, the resultant ``data-tsunami'' significantly increases the demand on Industry 4.0.
{\color{black}The types of information sources in Industry 4.0 are not limited to conventional symbols or bits, but include the underlying meaning or purpose to be delivered for industrial activities, which leads to the concept of {\it semantic communications}.}
In the face of such complex scenarios, more advanced physical layer techniques are required for wireless information transmission (WIT). Increasing the carrier frequency allows us to expand the bandwidth in the millimeter-wave (mmWave) and terahertz (THz) bands at the cost of increasing the path loss. Moreover, the use of high frequency bands facilitates the deployment of massive multiple-input multiple-output (MIMO) systems in a compact space, which can offer: 1) substantially improved beamforming gains to compensate for the increased path loss; and 2) fine-grained spatial multiplexing for simultaneously supporting multi-device communications.

The channels at higher frequencies are also vulnerable to blockage effects, hence the desired line-of-sight link is usually intermittent. This limitation led researches to the conception of a new paradigm for reconfiguring the wireless propagation environment via software-controlled devices for improving the channel conditions with the aid of reconfigurable intelligent surfaces (RISs) \cite{RIS}. In Industry 4.0 scenarios, inexpensive passive RISs can be deployed for enhancing the channel conditions to boost the system capacity. 

Another challenge is that of mitigating the effect of the high Doppler frequencies encountered by high-velocity robotic cars, mining vehicles, freight trains, and unmanned aerial vehicles (UAV) as well as planes. In these high-mobility scenarios, the pilot overhead used for channel estimation and equalization may become excessive, since every time the velocity is doubled, it also must be doubled. {In these scenarios, sophisticated non-coherently detected multiple-symbol receivers come to rescue}. Novel Doppler-resistant modulation schemes have also been proposed, such as, for example, orthogonal time frequency space (OTFS) modulation. Their applications in Industry 4.0 have to be further investigated.

\subsection{Heterogeneous Networks Management}
The networks in Industry 4.0 incorporate diverse components associated with different properties and requirements. Considering the three popular use cases of 5G networks as an example: 
\begin{itemize}
{\item {\it eMBB} requires abundant bandwidth for transmitting high-definition audio, pictures, videos, and even tactile tele-presence information.
}
{\item {\it mMTC}, which can be supported by LTE-M, NB-IoT, Lora, Sigfox, etc. requires high capacity to handle the simultaneous access requests arriving from a massive number of industrial devices. 
}
{\item {\it URLLC}, which can be supported both by 5G cellular networks and by dedicated short-range communications (DSRC), requires high reliability and low latency to support the precise control of machines.
}
\end{itemize}

This wide variety of specifications renders the networks in Industry 4.0 highly heterogeneous. Therefore it is quite a challenge to achieve efficient network management in these complex networking environments. The innovative concept of {\it network slicing} is potentially capable of coping with the challenges of heterogeneous networks,
{where the slices convey the different-specification information streams, as exemplified by critical control signalling, and best-effort Internet browsing}.
By combining the existing network technologies like virtual private networking, network function virtualization, software-defined networks, and so on, network slicing technology divides a networks into multiple segments, where each segment corresponds to different network requirements, and the segments do not affect each other. In this way, a network-slicing based architecture can simultaneously meet the different networking requirements of different scenarios, dealing with the heterogeneity of the network.

Intelligent AI-based network slicing reduces the cost of network management and improves service diversity \cite{Slicing}, which is appealing for Industry 4.0, but the provision of Quality of Service (QoS) and Quality of Experience (QoE) guarantees requires further research.

\begin{figure*}[!tp]
\centering
\captionsetup{singlelinecheck = off, font={footnotesize,color={black}}, name = {Fig.}, labelsep = period}
\subfigure{\includegraphics[width=5.5in]{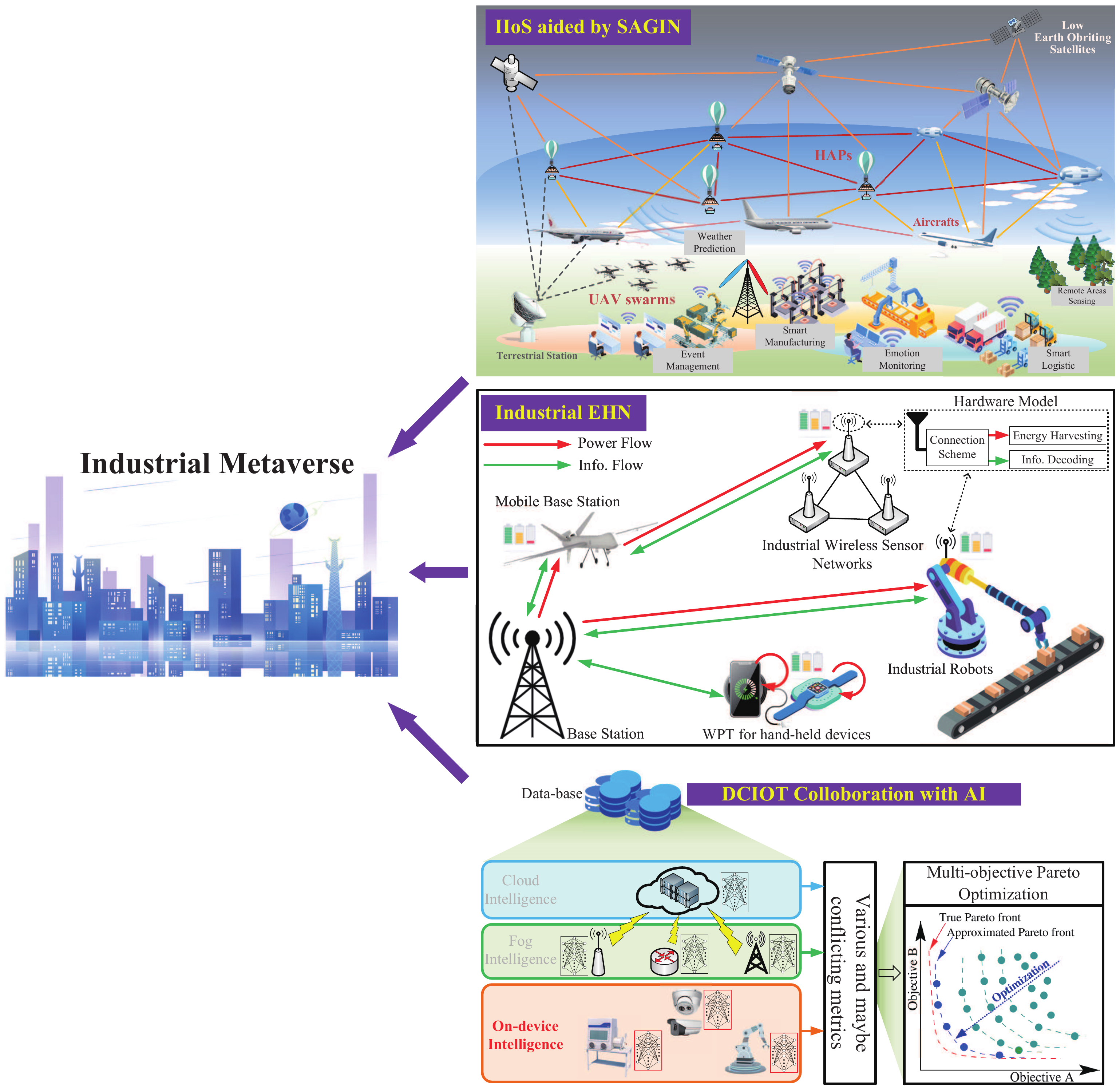}}
\caption{\color{black}Ambitious visions of Industry 4.0+.}
\label{F5}
\end{figure*}

\section{Future Visions, Opportunities, and Limitations}
In this section, we introduce some of the future visions and opportunities of Industry 4.0+, which are shown in the {stylized illustrations} of Fig.~\ref{F5}. Critical issues and limitations are discussed, which require further research.

{\color{black}
\subsection{Industrial Internet of Senses}
The paradigm shift from IIoT towards IIoS is promising for Industry 4.0+, which relies on the progress in the fields of both the ``Internet'' and in the communications of  ``Senses''.
As the space for human activities expands across the globe, the support for industry in vast and/or sparsely populated areas (space, oceans, forests, deserts, canyons, to name but a few) attracts much attention.
The interconnection, or ``Internet'', between the industrial elements in such scenarios is not feasible by simply relying on the existing terrestrial communication systems.
A key CT-domain enabler of fine-grained connectivity and high-accuracy positioning services is related to the space-air-ground integrated network (SAGIN) concept. By combining the terrestrial systems with the emerging OT-domain techniques, such as low Earth orbiting (LEO) satellites, high-altitude platforms (HAPs), aircraft, UAV swarms, etc., SAGIN will usher in a paradigm shift that will extend the breadth and depth of not only communication services, but also industrial activities.

Furthermore, the term ``Senses'' in the IoS encompasses a large range of concepts in the IT domains, which helps to support vast applications under Industry 4.0+, such as emission detection, weather prediction, event management, emotion monitoring for workers, etc.
In a nutshell, the IIoS assisted by SAGIN, as a typical form of the CIOT collaboration, is capable of monitoring the existing network environment through ubiquitous connection and real-time communications, thus simultaneously improving the QoS and QoE of future industrial networks.}

{\color{black}\subsection{Energy Harvesting Network}}
{\color{black}
%Charging a massive number of wireless elements in Industry 4.0+ using conventional wired charging methods is inconvenient, hence {power-beamforming or laser-charging} must be used.
%WPT is appealing for employment in Industry 4.0+, since it is inconvenient or potentially impossible to charge all wireless industrial elements, such as sensor nodes, via traditional wired links. 
Wireless power transfer (WPT) is an OT-domain technology that allows a power source to transmit electromagnetic energy to an IoT node over the air, without interconnecting wires \cite{WPT}.
Furthermore, WPT is waterproof or dustproof for contact-free devices, which is suitable for harsh industrial production environments in Industry 4.0+.
The current mechanisms of WPT can be categorized into the following types \cite{WPT}:
%WPT is appealing for employment in Industry 4.0+, since it is inconvenient or potentially impossible to charge all wireless industrial elements, such as sensor nodes, via traditional wired links. 

\begin{itemize}
{\item {\it RF charging,} which has a poor charging efficiency (about 1\%-10\%) but a remarkable charging distance (of the order of kilometers).
It can charge the remote devices operating in harsh environments, such as epidemic outbreak areas.
}
{\item {\it Inductive charging,} which guarantees more than 90\% charging efficiency, but only within a very small range (several centimeters).
It allows workers in the factories to expediently charge their hand-held devices.
}
{\item {\it Magnetic resonance based charging,} which strikes a charging efficiency vs. distance trade-off.
It constitutes the most promising type of WPT for sensor networks, electric vehicles, smart grid, etc. in Industry 4.0+.
}
\end{itemize}

More importantly, by combining WPT in the OT domain with WIT in the CT and IT domains, the concept of EHN, a.k.a., simultaneous wireless information and power transfer (SWIPT) \cite{WPT}, is established. With the rapid development of circuit design and implementation, industrial EHN can be readily realized with only minor hardware modification to industrial devices. By carefully designing the on-demand resource allocation scheme between the energy and information, EHN has the compelling advantage of concurrently delivering controllable and efficient wireless information and energy in industrial activities, hence SWPIT may be deemed more beneficial than the sum of its constituent parts (WPT and WIT).}

%\subsection{OT Evolution: Flexible Manufacturing}
%The operational industrial production lines fail to provide sufficient flexibility and reconfigurability for Industry 4.0.
%An example is found in aerospace manufacturing, which is a small-batch production activity creating very diverse components. It is impossible to build a dedicated digitalized production line for each type of component in practice. Therefore, diversifying the variety of components that can be produced efficiently becomes a critical issue for Industry 4.0 and beyond. However, both the risk and cost of reconfiguring the production lines should also be taken into account by human- and environment-friendly industries.
%Furthermore, reducing the reconfiguration time of production lines is also a challenge, especially for low-volume precious products.
%%but may require stringent standard and security (e.g., military weapons or frontier scientific instruments), it is necessary to build dedicated production lines in a timely and economical manner in order not to bungle the missions. This vision requires both cutting-edge manufacturing technology and management technology within a smart factory, therefore urges us to combine the efficiency of machine and the wisdom of human.
%%realizing customizable manufacturings in Industry 4.0 and beyond.

{\color{black}\subsection{Data-Driven CIOT Collaboration with AI}
Undoubtedly, AI will play, explicitly or implicitly, an important role for Industry 4.0 and Industry 4.0+, as well as for the industrial revolutions of the distant future.
Powerful AI techniques have been widely used in diverse research areas in the current era of machine intelligence for promoting the CIOT collaboration.
%However, one may ignore the importance of data technology (DT), represented by big data technique, during the application of AI.
%Expect for the CIOT domain, DT domain,  is also an indispensable part for future industrial network, which will become more and more essential.
Only by relying on a sufficiently large set of pre-collected data can AI tackle the model-free optimization problems in complex industrial networking setting.
Therefore, the CIOT collaboration may be further augmented with the aid of pervasive big data in the form of data-driven CIOT (DCIOT) collaboration, which holds the key for achieving high-level intelligence in Industry 4.0+.

Another critical challenge for AI-aided future industrial networks is their multi-objective optimization relying on multiple metrics.
Under the evolving paradigm of {\it on-device intelligence}, AI has been deployed in powerful IoT nodes, as well as in cloud centres (i.e., {\it cloud intelligence}) or smart APs and gateways (i.e., {\it fog intelligence}). This architecture is capable of improving both the latency and security, but results in conflicting requirements. One promising solution is based on the notion of multi-objective Pareto-optimization  \cite{Pareto} for finding all optimal operating points of a multi-component objective function constituting the optimal Pareto front. Suffice to say that for the solutions on the Pareto front, none of the system parameters can be improved without degrading at least one of the others.
In Industry 4.0+, this ubiquitous AI paradigm based on DCIOT collaboration and multi-objective Pareto optimization is capable of supporting self-configuring, self-optimizing, and self-adjusting production environments.}

\color{black}
\subsection{Industrial Metaverse}
The concept of the Metaverse is evolving rapidly, but its definition varies across different disciplines (see, e.g., \cite[Table I]{Metaverse}). In the context of industry, we define the industrial Metaverse as an industrial ecological system that intrinsically integrates DCIOT with the real-world industrial activities.
By harnessing the emerging techniques of XR, AI, IoT, cloud computing, blockchain, 5G/6G. etc., the industrial Metaverse supports the seamless connection of people and machines in an amalgamated network, acting as the vein of clean and intelligent manufacturing  as well as services. In this sense, the industrial Metaverse may be viewed as an evolutionary form of CIOT (or DCIOT) collaboration (see Fig.~\ref{F5}), relying on heterogeneous techniques and disciplines.

\subsection{The Limitations of This Study}
In this sub-section, we summarize some limitations of this study concerning the industrial revolution from a CIOT collaboration perspective, which are listed as follows.
\begin{itemize}
{\item For the manufacturing industries, only the most radical technologies that can be applied to the physical world have practical value. This may be engineered in from the
start for Industry 4.0+ to succeed.}
{\item As we argued, the availability of sufficient training data is a prerequisite for intelligent industry. However, much of the data are usually owned by private enterprises or governments, and they cannot be publicly exposed. Hence, in the face of limited data, intelligent industrial networking requires alternative solutions.}
{\item The multidisciplinary nature of industry requires the combination of expertise from different disciplines, including not only engineering, but also economics, sociology, and management, just to name a few. It is necessary to bridge the gap among the different disciplines in support of the evolution to Industry 4.0+.}
\end{itemize}
% Ref:https://baijiahao.baidu.com/s?id=1721531546555820783&wfr=spider&for=pc

\section{Conclusions}
Based on a CIOT collaboration perspective, this article has explored the landscape of Industry 4.0 and its beyond.
Specifically, the concept of CIOT collaboration has been introduced as a paradigm shift for future industry.
By seamlessly integrating the OTs found in the physical world with the ITs in the cyber world by relying on advanced CTs, the close CIOT collaboration is set to promote intensified industrial innovation.
The critical roles of the CT domain in the challenging use cases of Industry 4.0 have been highlighted.
On this basis, we have indicated some of research directions for the CT domain in terms of three salient aspects, namely the massive access paradigm , wireless information transmission, and network management.
Furthermore, we have extended our discussions to Industry 4.0+, where some ambitious visions are depicted by highlighting the interplay of the OT, IT, CT, and other disciplines.
The industrial revolution based on CIOT collaboration perspective is expected to reshape the industrial activities and transform our society.

\color{black}

\end{document}